\documentclass{aa}
\usepackage{amsmath}
\usepackage[varg]{txfonts}
\usepackage{graphicx,amssymb,float,subfigure}
\usepackage{natbib}
\bibpunct{(}{)}{;}{a}{}{,}

\begin{document}

\title{Thermonuclear supernova simulations with stochastic ignition}

\author{W. Schmidt\and J. C. Niemeyer}

\institute{Lehrstuhl f\"{u}r Astronomie, Institut f\"{u}r Theoretische Physik und Astrophysik, \\
  Universit\"{a}t W\"{u}rzburg, Am Hubland, D-97074 W\"{u}rzburg, Germany}

\date{Received / Accepted}

\titlerunning{Thermonuclear supernova simulations with stochastic ignition}
\authorrunning{W. Schmidt\and J. C. Niemeyer}

\abstract{We apply an \emph{ad hoc} model for dynamical ignition in
  three-dimensional numerical simulations of thermonuclear supernovae
  assuming pure deflagrations. The model makes use of the statistical
  description of temperature fluctuations in the pre-supernova core
  proposed by \citet{WunWoos04}.  Randomness in time is implemented by
  means of a Poisson process.  We are able to vary the explosion
  energy and nucleosynthesis depending on the free parameter of the
  model which controls the rapidity of the ignition process. However,
  beyond a certain threshold, the strength of the explosion saturates
  and the outcome appears to be robust with respect to number of
  ignitions. In the most energetic explosions, we find about
  $0.75M_{\sun}$ of iron group elements. Other than in simulations
  with simultaneous multi-spot ignition, the amount of unburned carbon
  and oxygen at radial velocities of a few $10^{3}\,\mathrm{km/s}$
  tends to be reduced for an ever increasing number of ignition
  events and, accordingly, more pronounced layering results.
	
}

\maketitle

\section{Introduction}

The significance of the ignition characteristics at the onset of
thermonuclear supernova explosions has been recognised since the
advent of multi-dimensional numerical simulations. The mounting
evidence for diversity among observed type Ia supernovae renders the
quest for realistic modelling of the ignition process yet more topical
\citep{BenCap05,LeoLi05,StrizLeib05}. Substantial efforts have been
made to analyse the effect of choosing different initial burning
regions in simulations of thermonuclear supernovae
\citep{GarBrav05,RoepHille05}. The multi-point ignition scenarios
which have recently been presented by \citet{RoepHille05c} suggest
that the energy generation becomes maximal for about one hundred
ignitions. In this case, unburned carbon and oxygen at low radial
velocity largely diminish.  Except for the two-dimensional study by
\citet{LivAsi05}, however, it has not been attempted so far to
generate ignitions at different instants of time during a simulation.

In this article, we present a simple stochastic model for the
dynamical ignition of thermonuclear deflagration in three-dimensional
simulations.  The purpose of the model is an investigation of the
possible consequences of using a simple parametrisation for the
ignition probability. In particular, the total amount of iron group
and intermediate mass elements as well as their spatial distribution
and density functions in radial velocity space are studied. We do not
claim yet that the highly complex physics of the ignition process in
thermonuclear supernova is fully taken into account. Notwithstanding
the tentative nature of the stochastic ignition model, it may help to
clarify the extent of possible variations in the yields and the
distribution of burning products without invoking a hypothetical
deflagration to detonation transition \citep{GamKhok04,GamKhok05}. Whether
the predictions of the model can be reconciled with properties of
observed type Ia supernovae (SNe Ia) will depend on further
developments in the theoretical understanding of convection and the
birth of flames in the pre-supernova core.

\section{Stochastic ignition}

The simplest thermodynamical description of the pre-supernova core is
an adiabatic background plus temperature fluctuations due to
convection driven by gradual thermonuclear burning. For the background
temperature, we assume the parabolic profile
\begin{equation}
	T_{\mathrm{a}}(r) = \max\left\{T_{0}\left[1-\left(\frac{r}{\Lambda}\right)^{2}\right], 
	                              T_{\mathrm{ext}}\right\}
\end{equation}
with a cutoff at the thermal radius $\Lambda=7.35\cdot
10^{7}\,\mathrm{cm}$ for the central density $\rho_{\mathrm{c}}=2\cdot
10^{9}\,\mathrm{g\,cm^{-3}}$ \citep{WunWoos04} and an isothermal
exterior region with $T_{\mathrm{ext}}=5\cdot 10^{5}\,\mathrm{K}$. For
the temperature fluctuations, \citet{WunWoos04} proposed a statistical
model on the basis of the mixing-length approach. The typical
temperature excess $\delta T_{\mathrm{b}}$ gained by fluid elements
relative to the adiabatic background in a spherically symmetric core
region with Kolmogorov power spectrum is given by
\begin{equation}
  \frac{\delta T_{\mathrm{b}}}{T_{0}}\sim
  \left(\frac{\Lambda}{R}\right)^{4/7}\left(G\rho_{0}\delta_{P}\tau^{2}\right)^{-3/7},
\end{equation}
where $R$ can be interpreted as mixing length, $\delta_{P}$ is the
logarithmic derivative of density relative to pressure and $\tau$ is
the nuclear time scale \citep[see][ eqn.~9 \& 18]{WunWoos04}.

\begin{table*}[htb]
  \caption{Parameter study with initial spatial resolution $\Delta = 4.75\,\mathrm{km}$
     in the uniform part of the numerical rid. The total spherical angle covered
     by the simulation domain is $\Omega$, the total number of
     ignitions per octant is $I_{\pi/2}$,
     and the last ignition event occurs at time $t_{\mathrm{last}}$.}
  \label{tb:simulations}
  \begin{center}
    \begin{tabular}{r l c r c c c c c c}
      \hline
      $C_{\mathrm{e}}$ & $R/\Lambda$ & $\Omega$ & $I_{\pi/2}$ & $t_{\mathrm{last}}\,[s]$ & 
      $E_{\mathrm{nuc}}\,[10^{51}\,\mathrm{erg}]$ & $E_{\mathrm{kin}}\,[10^{51}\,\mathrm{erg}]$ & 
      $M_{\mathrm{Ni}}/M_{\sun}$ & $M_{\mathrm{Mg}}/M_{\sun}$ \\
      \hline\hline
      $10^{1}$        & 0.5  & $\pi/2$ &   2 & 0.200 & 0.516 & 0.008 & 0.242 & 0.165 \\
      $10^{2}$        & 0.5  & $\pi/2$ &   4 & 0.096 & 0.827 & 0.320 & 0.413 & 0.218 \\
      $10^{3}$        & 0.5  & $\pi/2$ &  30 & 0.508 & 1.139 & 0.631 & 0.629 & 0.187 \\
      $5\cdot 10^{3}$ & 0.5  & $\pi/2$ &  65 & 0.177 & 1.219 & 0.711 & 0.690 & 0.170 \\
      $10^{4}$        & 0.25 & $\pi/2$ &  65 & 0.199 & 1.099 & 0.591 & 0.603 & 0.189 \\
      $10^{4}$        & 0.5  & $\pi/2$ & 115 & 0.323 & 1.261 & 0.753 & 0.707 & 0.188 \\
      $10^{4}$        & 1.0  & $\pi/2$ & 190 & 0.203 & 1.157 & 0.649 & 0.643 & 0.184 \\
      $2\cdot 10^{4}$ & 0.5  & $\pi/2$ & 131 & 0.222 & 1.162 & 0.654 & 0.643 & 0.188 \\
      $10^{5}$        & 0.5  & $\pi/2$ & 284 & 0.306 & 1.131 & 0.623 & 0.629 & 0.177 \\
      $10^{2}$        & 0.5  & $4\pi$  &   6 & 0.703 & 0.985 & 0.478 & 0.523 & 0.201 \\
      $5\cdot 10^{3}$ & 0.5  & $4\pi$  &  59 & 0.237 & 1.274 & 0.766 & 0.715 & 0.188 \\
      $10^{5}$        & 0.5  & $4\pi$  & 305 & 0.159 & 1.165 & 0.657 & 0.647 & 0.184 \\
      \hline
    \end{tabular}
  \end{center}
\end{table*}

The ignition algorithm utilised in our simulations is based on a
discrete stochastic process, namely, the Poisson process.  The
probability for $i$ ignitions within the time interval
$\delta t$ is given by
\begin{equation}
  \label{eq:poisson_prob}
  P_{t,\vec{r}}(i,\delta t)=
  \frac{(\nu_{t,\vec{r}}\delta t)^{i}}{i!}\exp(-\nu_{t,\vec{r}}\delta t),
\end{equation}
where $\nu_{t,\vec{r}}$ is a function of the temperature excess $\delta T_{\mathrm{b}}$
and the local thermodynamic background state.  
Assuming that the ignition events are statistically independent, the
number of ignitions per unit time increases proportional to the volume
of fuel within a region of approximately constant background conditions.
For this reason, we consider the ignition probability
$P_{t,r,\delta r}(i,\delta t)$ for the unburned material within
spherical shells of width $\delta r\ll R$. On grounds of
ergodicity, the increase of probability in proportion to the volume is
equivalent to either increasing the sampling time $\delta t$ or
decreasing $\nu_{t,\vec{r}}$ by the ratio of the fuel volume relative
to a unit volume. Thus, $P_{t,r,\delta r}(i,\delta t)$ is given by an
expression analogous to the right hand side in
equation~(\ref{eq:poisson_prob}) with $\nu_{t,\vec{r}}$ replaced by
\begin{equation}
  \nu_{t,r,\delta r}=
  \langle\nu_{t,\vec{r}}\theta[G(t,\vec{r})]\rangle_{r,\delta r}
  \frac{3r^{2}\delta r}{r_{\mathrm{b}}^{3}},
\end{equation}
where $G(t,\vec{r})$ is the level set function which represents the
flame front, $\theta[G(t,\vec{r})]$ is zero for $G(t,\vec{r})<0$
(i.e.~in ash regions) and unity for $G(t,\vec{r})>0$. $\langle\
\rangle_{r,\delta r}$ denotes the average over a spherical shell of
radius $r$ and width $\delta r$. For
the normalisation, we choose a sphere of radius $r_{\mathrm{b}}$ which
is roughly the size of the central heating region \citep[][
eqn.~24]{WunWoos04}. The underlying assumption of spherical symmetry
is, of course, increasingly flawed in the course of the
explosion. However, we found empirically that the spatial variation of
the background state variables within the unburned material contained
in a thin spherical shell at each given instant of time is small. 

In order to complete the model, we formulate a hypothesis for
$\langle\nu_{t,\vec{r}}\theta[G(t,\vec{r})]\rangle_{r,\delta
r}$.  Since $\nu_{t,\vec{r}}$ is an inverse time scale which
determines the probability of ignition per unit time, we conjecture
that
\begin{equation}
  \label{eq:poisson_nu}
  \chi_{\mathrm{u},r,\delta r}(t)^{-1}\langle\nu_{t,\vec{r}}\theta[G(t,\vec{r})]\rangle_{r,\delta r}=
  C_{\mathrm{e}}\frac{S_{\mathrm{nuc}}}{c_{P}\delta T_{\mathrm{b}}},
\end{equation}
where $c_{P}$ is the heat capacity and $S_{\mathrm{nuc}}$ is the
nuclear energy generation rate for the average temperature and mass
density, respectively, of unburned material within a particular
spherical shell \citep[][ eqn.~4 and~6]{WoosWun04}. The exponentiation
parameter $C_{\mathrm{e}} $ determines the overall rapidity of the
ignition process. The filling factor $\chi_{\mathrm{u},r,\delta r}(t)$
specifies the volume fraction of unburned fluid at time $t$ within the
shell. Note that both sides of equation~(\ref{eq:poisson_nu}) are
independent of the amount of fuel which is consistent with the
statistical nature of the quantities involved.

\begin{figure*}[thb]
  \begin{center}
    \includegraphics{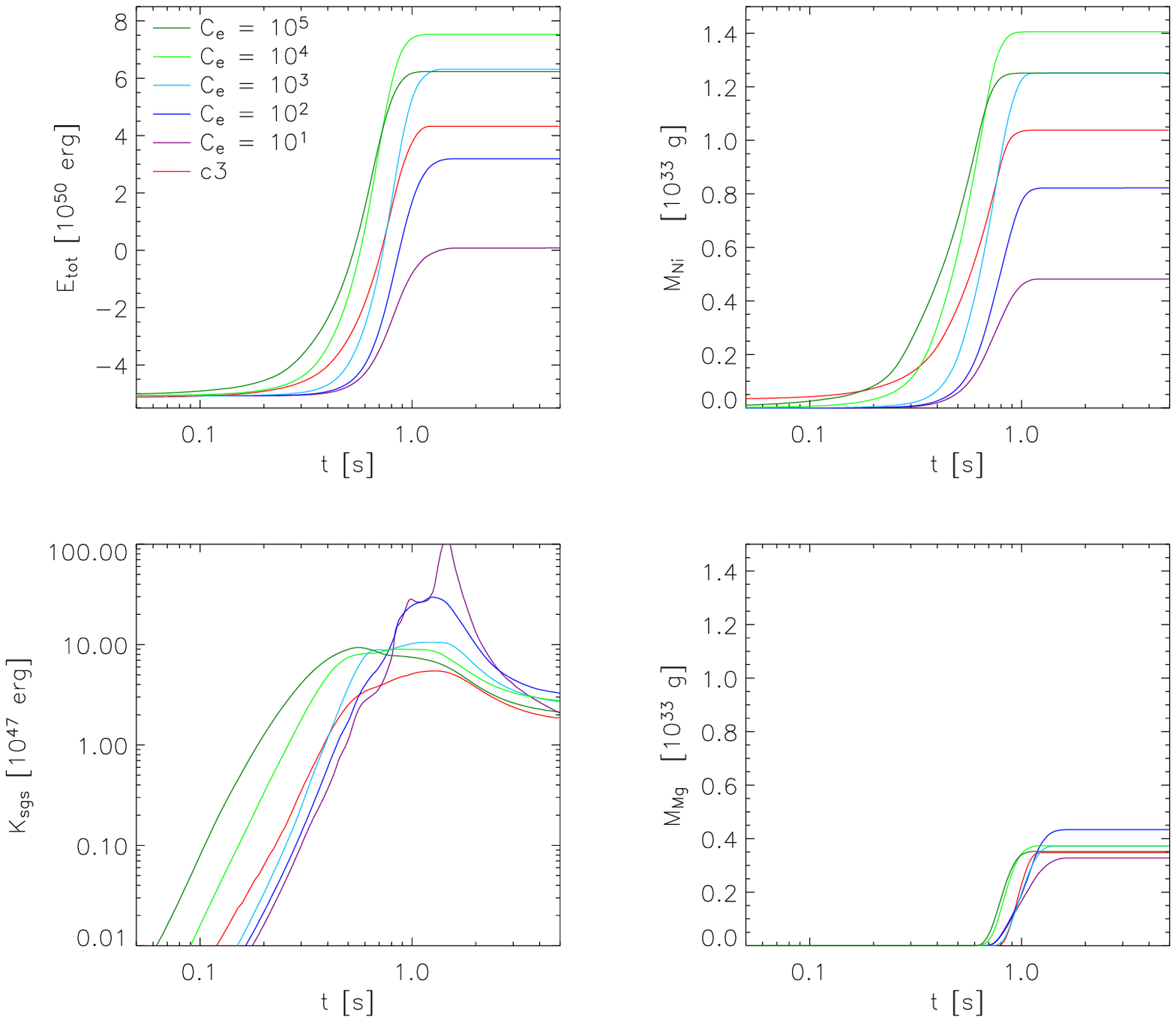}
    \caption{Evolution of the total energy, the masses of iron group
      and intermediate mass elements, respectively, as well as the
      subgrid scale turbulence energy in simulations with axisymmetric
      initial flame (c3) and stochastic ignition with varying
      exponentiation parameter $C_{\mathrm{e}}$, respectively. The
      computational domain is a single octant with $N=128^{3}$ grid
      cells.  }
    \label{fg:evol_energy}
  \end{center}
\end{figure*}

\section{Numerical simulations}

For the numerical simulations presented in the following, we use the
same techniques as described in \citet{RoepHille05}, except for a
hybrid grid geometry with a central uniform part and exponentially
increasing grid size outside \citep{RoepHille05c} in combination with
a localised subgrid scale model which is introduced in
\citet{SchmHille05}. Owing to the co-moving hybrid grid, we were able
to achieve sensible results with merely $128^{3}$ numerical cells in
one octant. The subgrid scale model is based on the hydrodynamical
equation for the turbulence energy which was adopted by
\citet{NieHille95} for the numerical simulation of thermonuclear
supernovae.  In the new implementation, we apply dynamical procedures
for the \emph{in situ} calculation of closure parameters. Moreover, an
additional term was added to the subgrid scale turbulence energy
equation which accounts for buoyancy effects on unresolved scales
\citep{SchmNie05}. Thereby, the Sharp-Wheeler relation for the
characteristic velocity scale of the Rayleigh-Taylor instability is
effectively incorporated into the turbulence energy model.  In
addition, the localised SGS model features an algorithm which corrects
the SGS turbulence energy for the effect of the shifting cutoff length
scale of the co-moving grid, because the gradually increasing
cutoff length entails an intrinsic growth of SGS turbulence energy
\citep{SchmNie05}. 

The numerical implementation of the Poisson process was adopted from
\citet{NumRecip}. For each ignition event in a particular Poisson
process realisation, a random point is sampled from a uniform spatial
distribution and projected onto the spherical shell associated with
the process. If the resulting point happens to fall inside fuel, a
small bubble of burned material with radius $1.5\Delta$, where
$\Delta$ is the numerical cell size, is inserted. This algorithm
implicitly accounts for the filling factor $\chi_{\mathrm{u},r,\delta
r}(t)$ of the fuel in each shell.  For the width of the spherical
shells we chose $\delta r=2\Delta$.

\begin{figure}[thb]
  \begin{center}
    \resizebox{\hsize}{!}{\includegraphics{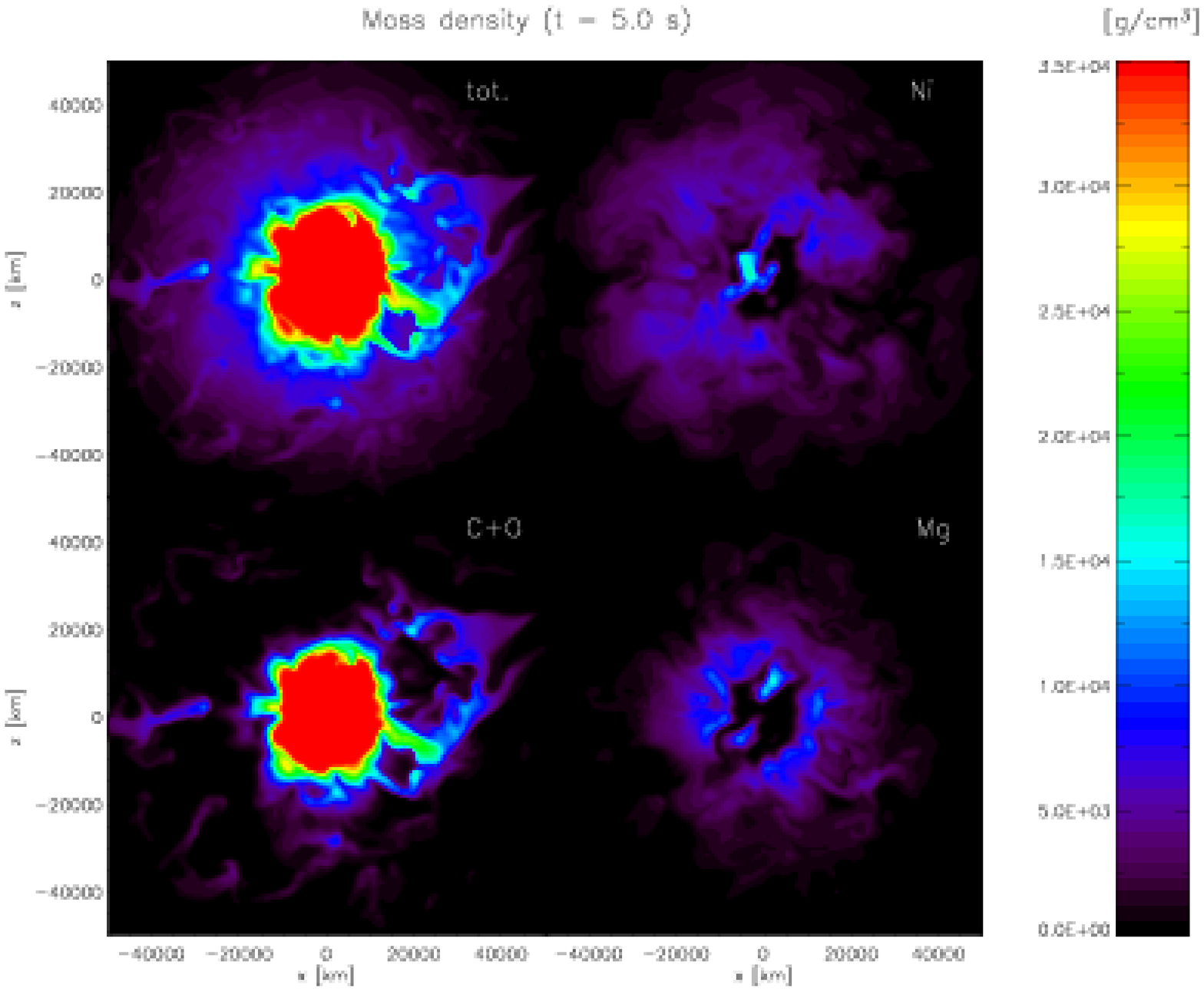}}
    \resizebox{\hsize}{!}{\includegraphics{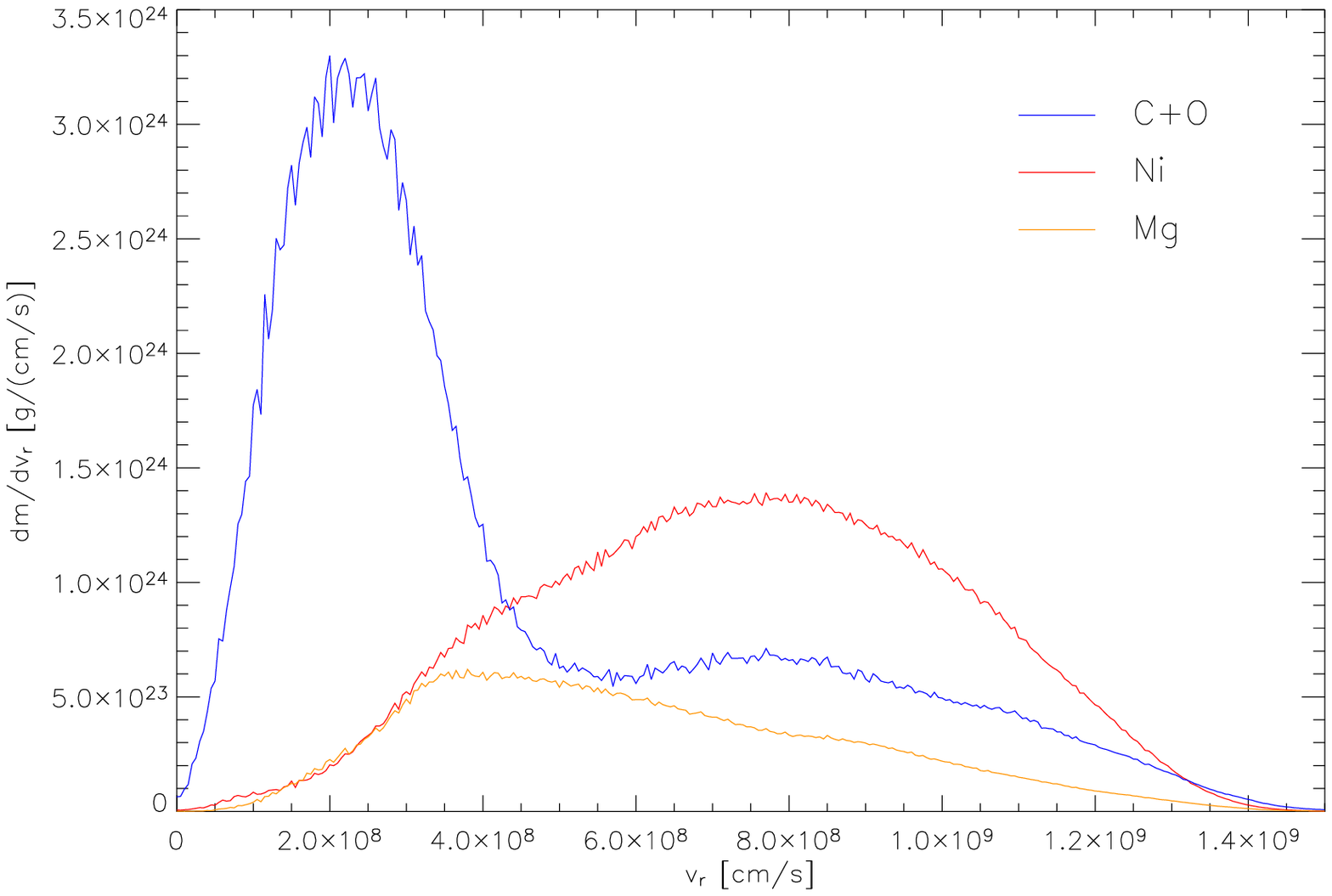}}
    \caption{Total and fractional mass densities in a central
      section and the corresponding probability density functions
      in radial velocity space for a full-star simulation with
      $C_{\mathrm{e}}=10^{2}$ at $t=5.0\,\mathrm{s}$.
    }
    \label{fg:4pi_256i}
  \end{center}
\end{figure}

\begin{figure}[thb]
  \begin{center}
    \resizebox{\hsize}{!}{\includegraphics{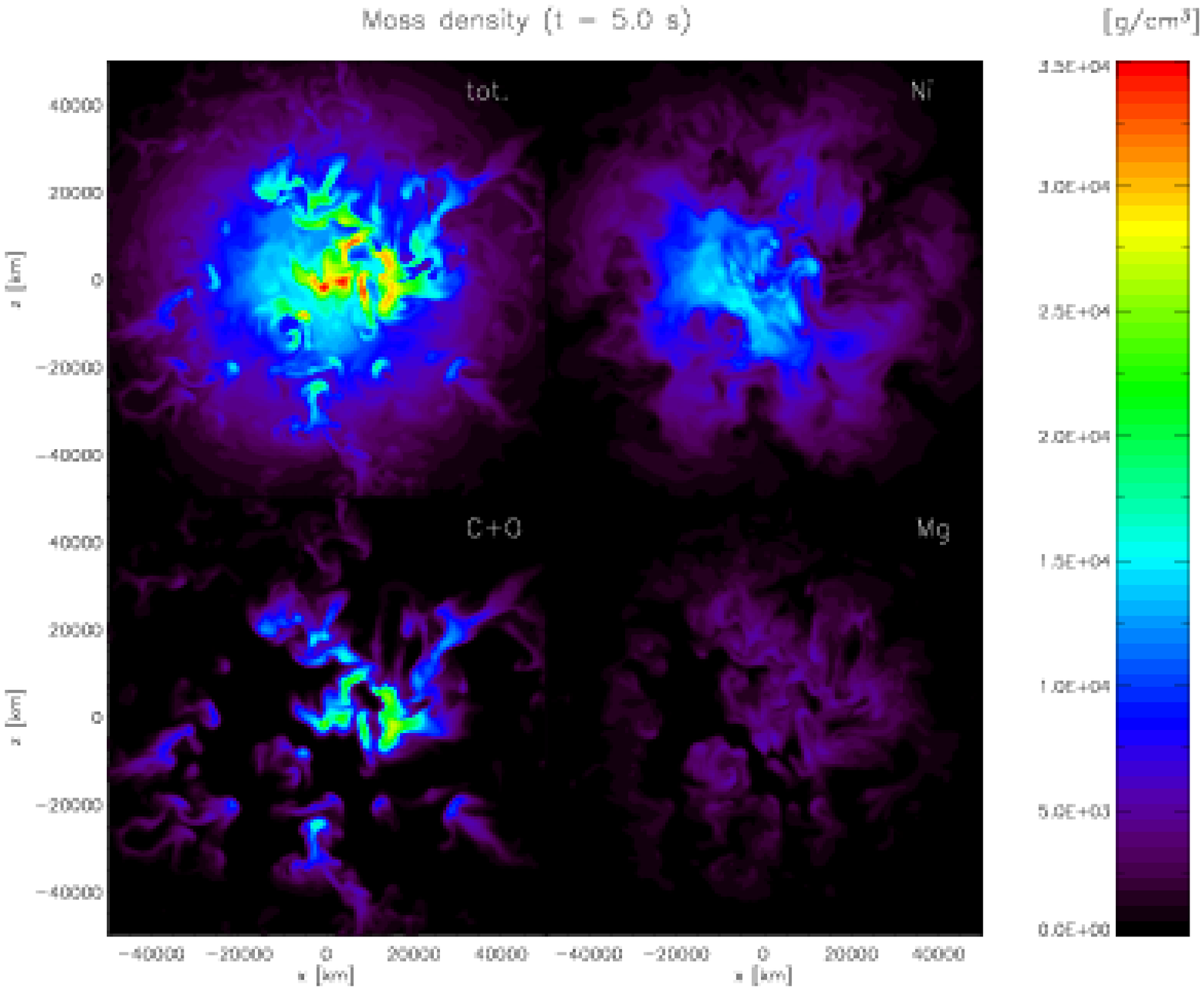}}
    \resizebox{\hsize}{!}{\includegraphics{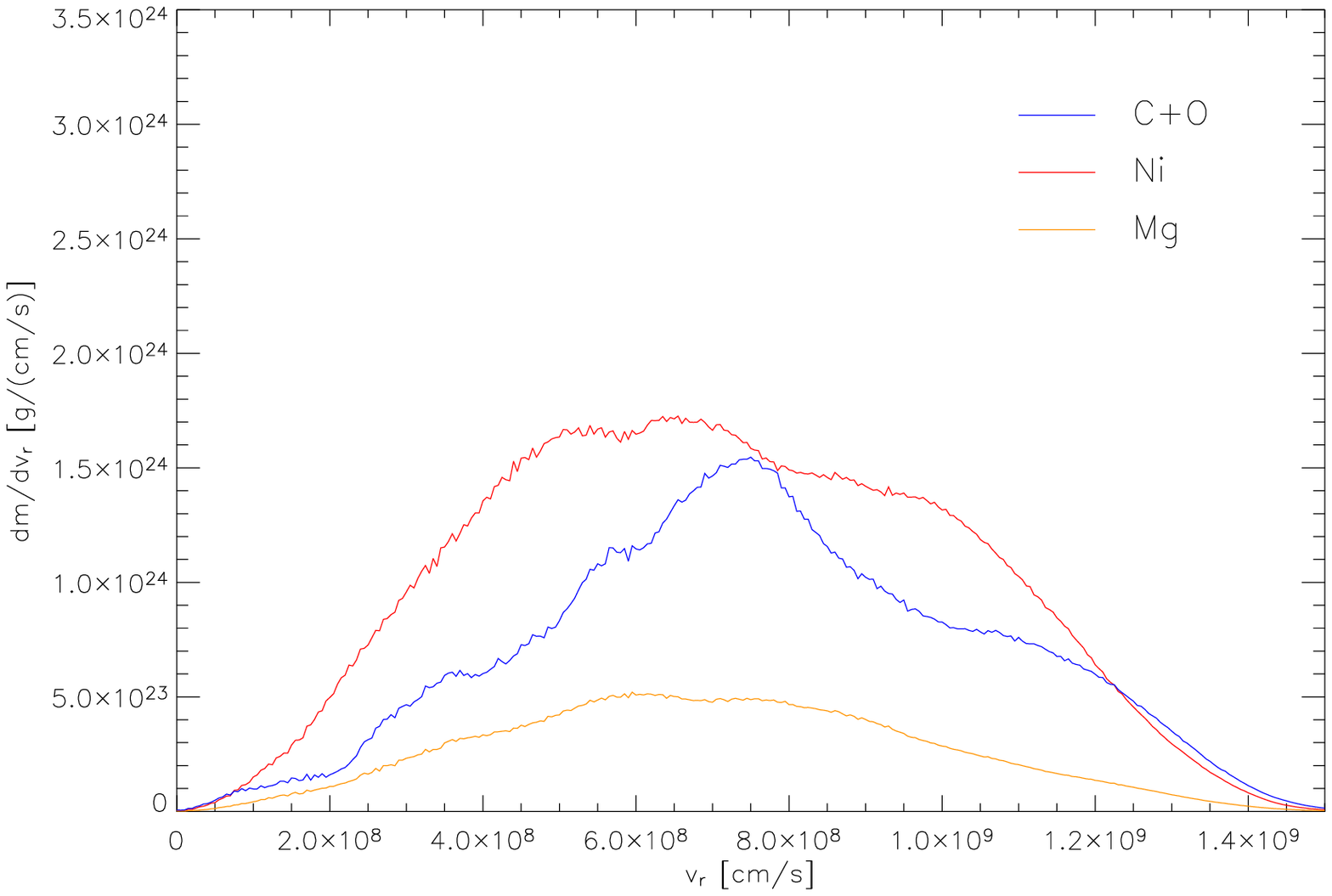}}
    \caption{Total and fractional mass densities in a central
      section and the corresponding probability density functions
      in radial velocity space for a full-star simulation with
      $C_{\mathrm{e}}=5\cdot 10^{3}$ at $t=5.0\,\mathrm{s}$.
    }
    \label{fg:4pi_256ii}
  \end{center}
\end{figure}

\begin{figure}[thb]
  \begin{center}
    \resizebox{\hsize}{!}{\includegraphics{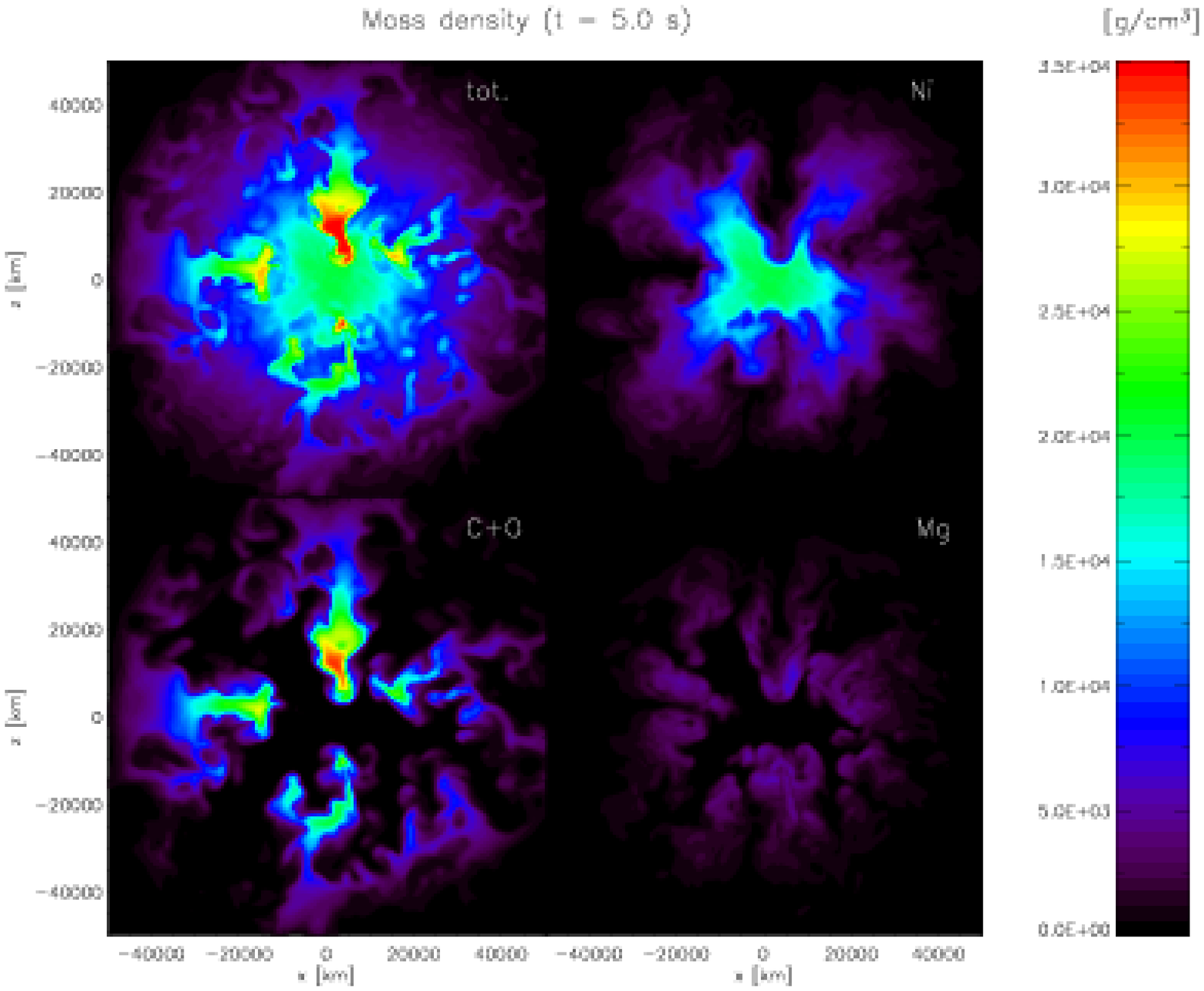}}
    \resizebox{\hsize}{!}{\includegraphics{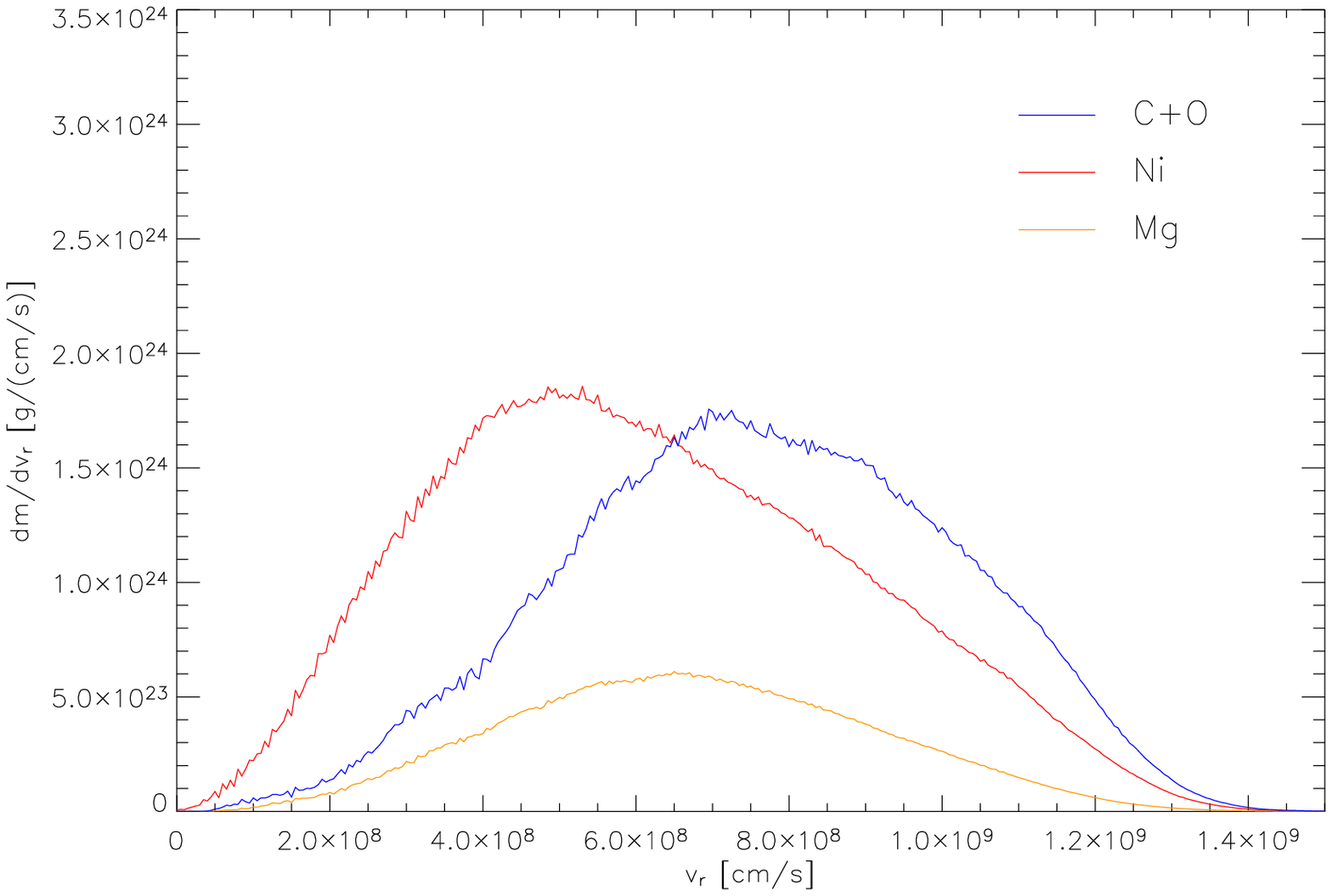}}
    \caption{Total and fractional mass densities in a central
      section and the corresponding probability density functions
      in radial velocity space for a full-star simulation with
      $C_{\mathrm{e}}=10^{5}$ at $t=5.0\,\mathrm{s}$.
    }
    \label{fg:4pi_256iii}
  \end{center}
\end{figure}

We performed simulations with different values of $C_{\mathrm{e}}$
both in octants and full spherical domains. The influence of different
mixing lengths $R$ relative to the thermal radius $\Lambda$ was
investigated as well.  The simulation parameters as well as the total
number of ignitions, the explosion energetics and the burning product
masses are listed in Table~\ref{tb:simulations}. In one series of
simulations the value of the exponentiation parameter $C_{\mathrm{e}}$
was varied over four orders of magnitude. The main conclusion is that
the explosion energy and the total mass of iron group elements
(represented by ${^{56}}$Ni) is maximal for $C_{\mathrm{e}}\sim
10^{4}$.  If the ignition process becomes even more rapid, bubbles of
burning material are generated at such high rate that most of the
bubbles merge before they can appreciably expand and deform due to the
Rayleigh-Taylor instability. This effect tends to inhibit the growth
of the flame surface area and, thus, reduces the explosion
energy. More or less the same trend was observed for simultaneous
ignition with varying bubble number by \citet{RoepHille05c}.

For $C_{\mathrm{e}}\ll 10^{3}$, there are only few ignitions per
octant.  In this case, the explosion is weak and the total mass of
iron group elements is less than $0.5M_{\sun}$. Setting
$C_{\mathrm{e}}=10$, the white dwarf becomes barely
unbound. Nevertheless, the ignition process ceases because the
probability of further ignitions becomes virtually zero once the star
is expanding appreciably. This is simply a consequence of the
power-law dependence of the factor $S_{\mathrm{nuc}}/c_{P}$ in
formula~(\ref{eq:poisson_nu}) on the background temperature and
density.  In general, the ignition process terminates after a few
tenths of a second regardless of the exponentiation parameter (see
Table~\ref{tb:simulations}). The case $C_{\mathrm{e}}=1$ corresponds
to roughly one ignition per second and would fail to produce an
explosion at all. On the other hand, there appears to be a remarkable
robustness for $C_{\mathrm{e}}\gtrsim 10^{3}$. The variation in the
total nuclear energy release is less than $10\,\%$ for all simulations
with exponentiation parameter in excess of $10^{3}$, while the number
of ignition events may change by a factor as large as ten.  Changing
the ratio of the mixing length to the thermal radius has more or less
the same effect as varying the exponentiation parameter. If
$R/\Lambda$ increases, the probability of ignition becomes
higher. However, we found that choosing $R/\Lambda$ significantly
smaller than $0.5$ introduces an artificial cutoff in the range of
ignition radii. For this reason, we set $R/\Lambda=0.5$.

The time evolution of the total energy is plotted for selected
simulations in panel (a) of Fig.~\ref{fg:evol_energy}. One can clearly
see the faster progression of the explosion with increasing number of
ignitions and the reduced rate of growth for $C_{\mathrm{e}}\sim
10^{5}$.  The final kinetic energy of $7.5\cdot 10^{50}\,\mathrm{erg}$
in the case $C_{\mathrm{e}}\sim 10^{4}$ is about $75\,\%$ larger than
for the reference simulation with a perturbed axisymmetric initial
flame (c3).  The production of iron group and intermediate mass
elements is shown in panels (b) and (d), respectively. It becomes
clear that the statistical properties of the ignition process are
crucial for the nucleosynthesis of iron group elements, whereas the
the final amount of intermediate mass elements, which are represented
by ${^{24}}$Mg, appears to be fairly robust. However, it is important
to note that burning is terminated at the density threshold of
$10^{7}\,\mathrm{g\,cm^{-3}}$ because we still lack a sensible
treatment of the distributed burning regime during the late stage of
the explosion. Continuing the burning process into this regime might
entail significant changes in the predictions of intermediate mass
element production.

The integrated subgrid scale turbulence energy $K_{\mathrm{sgs}}$ is
plotted in panel (c) of Fig.~\ref{fg:evol_energy}. In the early phase
of the explosion, one can discern power-law growth of
$K_{\mathrm{sgs}}$, where the exponent appears to be universal for
stochastic ignition. There is merely a temporal shift of the graph
depending on the exponentiation parameter. From $t\approx
0.5\,\mathrm{s}$ onwards, the increase of SGS turbulence energy
diminishes. This marks the onset of the fully turbulent regime. The
peak of $K_{\mathrm{sgs}}$ is increasingly delayed as the ignition
process progresses slower for smaller values of $C_{\mathrm{e}}$.  At
still later time, the turbulence energy decreases due to the rapid
bulk expansion and the quenching of the burning process which supplies
energy to the flow. The exceptionally high peak values of
$K_{\mathrm{sgs}}$ for the less energetic explosions are caused by a
major fraction of the released energy stirring the white dwarf matter into
turbulent motion rather than being consumed by bulk expansion.  In
this respect, the ``dud explosions'' obtained in the case of slow
ingition are more like an overflowing boiling pot.

\begin{figure}[thb]
  \begin{center}
    \resizebox{\hsize}{!}{\includegraphics{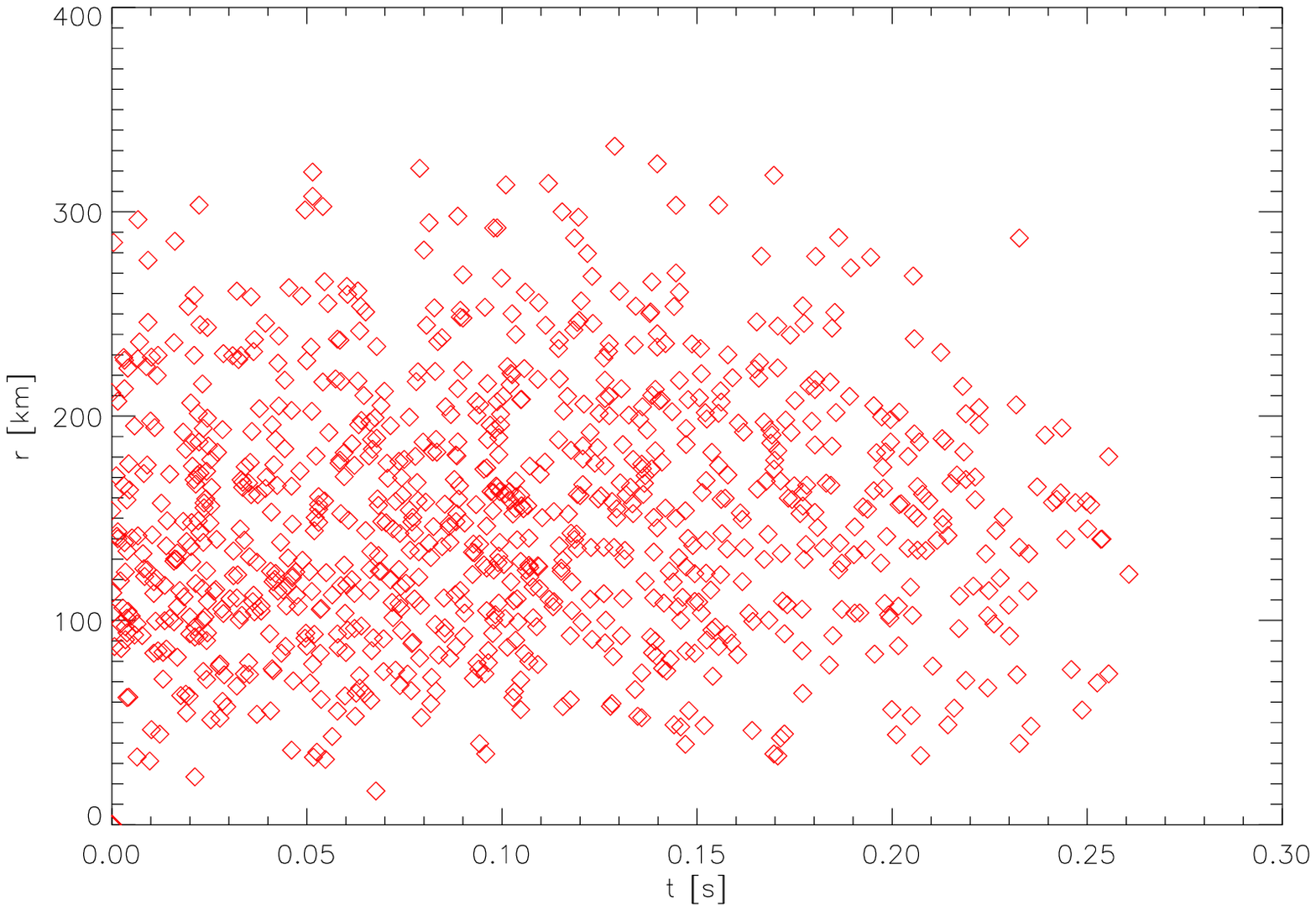}}
    \caption{Plot of the radii of ignition events as function of time in the
      full-star simulation with $N=512^{3}$.
    }
    \label{fg:ignt}
  \end{center}
\end{figure}

With regard to the question whether deflagration models can account
for observed type Ia supernovae, it is crucial to look at the
distribution of nuclear species at a time when the explosion ejecta
enter the phase of homologous expansion. To that end we performed
three full-star simulations and analysed the ejecta at
$t=5.0\,\mathrm{s}$. From the statistics listed in
Table~\ref{tb:simulations} it might appear that the sensitivity to the
ignition process is less pronounced than for the single-octant
simulations, possibly due to the absence of artifical boundary
conditions. However, Fig.~\ref{fg:4pi_256i}--\ref{fg:4pi_256iii}
illustrate that there are, indeed, substantial differences in the
final outcome.  For each simulation, the stratification of the mass
density is shown in two-dimensional sections through the centre of the
computational domain. It is instructive to compare the partial
densities of unburned carbon and oxygen, iron group elements (``Ni'')
and intermediate mass elements (``Mg''). In the case
$C_{\mathrm{e}}=10^{2}$, the contour sections on the top of
Fig.~\ref{fg:4pi_256i} show that a big lump of unburned material prevails
in the central region. The corresponding fractional masses as function of radial
velocity are plotted at the bottom of Fig.~\ref{fg:4pi_256i}. Given
the total amount of roughly $0.5\,M_{\sun}$ of iron group
elements, such an explosion definitely does not resemble any observed
SN Ia.  Increasing the exponentiation parameter, however, we obtain an
ever more stratified distribution of nuclear species. For
$C_{\mathrm{e}}=5\cdot 10^{3}$, the nuclear ash appears to be mixed
with unburned material over a wide range of radii and radial
velocities, respectively (Fig.~\ref{fg:4pi_256ii}). In the case of
most rapid ignition with $C_{\mathrm{e}}=10^{5}$, a clear trend
towards concentration of nickel in the centre becomes apparent, while
carbon and oxygen remain mostly in the outer layers.  This is
different from the results reported by \citet{RoepHille05c} for
simultaneous ignitions, where the material at small radial velocities
appears to be depleted of carbon and oxygen for a total number of
initially burning bubbles of the order one hundred, while further
increasing the number of bubbles reverses the trend.

For a more detailed account of the flame evolution, we performed in
addition one full-star simulation with the intermediate exponentiation
parameter $C_{\mathrm{e}}=5\cdot 10^{3}$ using $N=512^{3}$ numerical
cells, i.e. twice the initial spatial resolution. The ignition process
for this simulation is illustrated in Fig.~\ref{fg:ignt}. It might
seem unexpected that relatively few ignitions occur close to the
centre.  However, this is a natural consequence of the scaling of the
ignition probability with the volume. Moreover, most of the material
near the centre is burned quickly and so there remains little space
for further ignitions. The bulk of ignition events takes place at
radii in the range between $50$ and $250\,\mathrm{km}$ from the
centre.  The relatively large radial spread might be an artifact of
the statistical rather than local evaluation of the temperature excess
$\delta T_{\mathrm{b}}$ in expression~(\ref{eq:poisson_nu}) for the
inverse time scale of the Poisson process. In consequence, we
definitely see the requirement of an improved modelling of the
temperature fluctuations. The conceptual design of a fluctuation model
which allows for the computation of the spatiotemporal distribution
has recently been proposed by \citet{Kerst05}.

Three-dimensional snapshots of the flame fronts in the simulation
with high resolution are shown in Fig.~\ref{fg:pop_4pi}. The subgrid
scale turbulence velocity which determines the turbulent flame speed
is indicated by contours at the flame surface. At time
$t=0.1\,\mathrm{s}$ (a), one can see a large number of small bubbles
generated by the stochastic ignition process. As the burning bubbles are
floating outwards from the centre they begin to form the typical
Rayleigh-Taylor mushroom shapes (b).  At $t=0.5\,\mathrm{s}$, the flames
mostly have merged into a single structure which is subject to
intense turbulent flow (c). Apart from the small-scaling wrinkling
of the flame front one can discern several dominating modes. Particularly,
a pronounced asphericity emerges after most of the explosive burning
has come to pass (d). The formation of dominating modes is a consequence
of the fast growth of Rayleigh-Taylor instabilities seeded at
large radii. Hence, the evolution of the flames is highly sensitive
to the radial distribution of the ignitions. This implication
is particularly interesting in the light of recently found
observational evidence for asphericity of type Ia supernova ejecta
\citep{LeoLi05}. The late-time evolution and, particularly, the
distribution of nuclear species was found to be quite similar
to the $256^{3}$ run.

\section{Conclusion}

We have achieved a first step toward the dynamical modelling of the
ignition process in three-dimensional large-scale simulations of
thermonuclear supernovae. There is one free parameter, called the
exponentiation parameter $C_{\mathrm{e}}$, which controls the
statistical properties of the ignition process, especially, the total
number of ignition events.  Varying $C_{\mathrm{e}}$ in the numerical
simulations alters the total mass of iron group elements and the
stratification of burning products.  Particular ignition events can
affect the morphology of the explosion ejecta as some modes grow
faster than others. In fact, we found noticeable asphericity of the
developing flame front even for high values of $C_{\mathrm{e}}$, where
the statistical distribution of ignitions is nearly spherically
symmetric.

If the total number of ignitions is less than one hundred in the whole
core, only weak explosions with small total masses of iron group
elements result. Although certain type Ia supernovae appear to
produce, for instance, as little as $0.1M_{\sun}$ of nickel
\citep{StrizLeib05}, the exceptionally large amount of left-over
carbon and oxygen found in the simulations with small number of
ignitions renders them unlikely to correspond to real events.  The
picture might change, however, if burning at low density is properly
taken into account and, possibly, substantially larger amounts of
intermediate mass elements are produced.

\begin{figure*}[thb]
  \begin{center}
    \mbox{\subfigure[$t=0.1\,\mathrm{s}$, $N_{\mathrm{uni}}=320^{3}$,
          $X_{\mathrm{uni}}=815\,\mathrm{km}$]{\includegraphics[width=7.5cm]{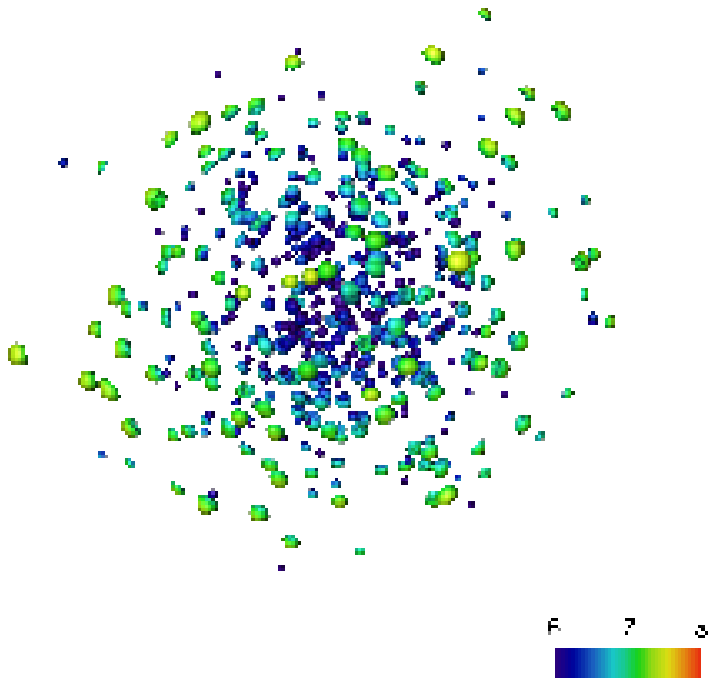}}\quad
          \subfigure[$t=0.25\,\mathrm{s}$, $N_{\mathrm{uni}}=340^{3}$,
          $X_{\mathrm{uni}}=1125\,\mathrm{km}$]{\includegraphics[width=7.5cm]{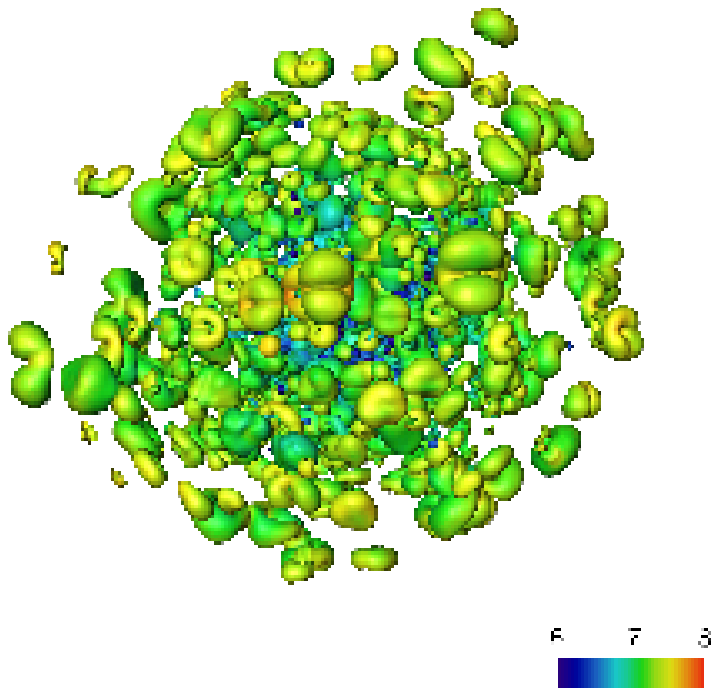}} }
    \mbox{\subfigure[$t=0.5\,\mathrm{s}$, $N_{\mathrm{uni}}=396^{3}$,
          $X_{\mathrm{uni}}=2339\,\mathrm{km}$]{\includegraphics[width=7.5cm]{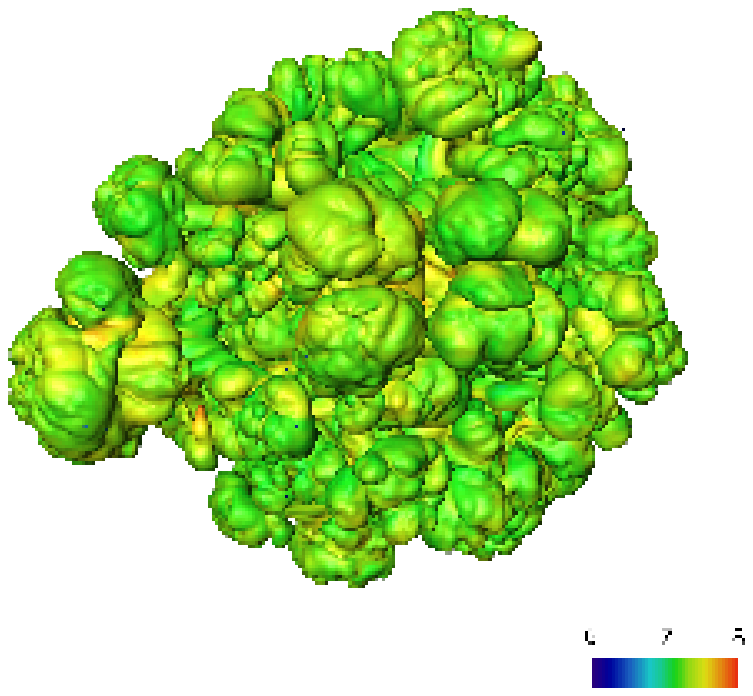}}\quad
          \subfigure[$t=0.75\,\mathrm{s}$, $N_{\mathrm{uni}}=438^{3}$,
          $X_{\mathrm{uni}}=5177\,\mathrm{km}$]{\includegraphics[width=7.5cm]{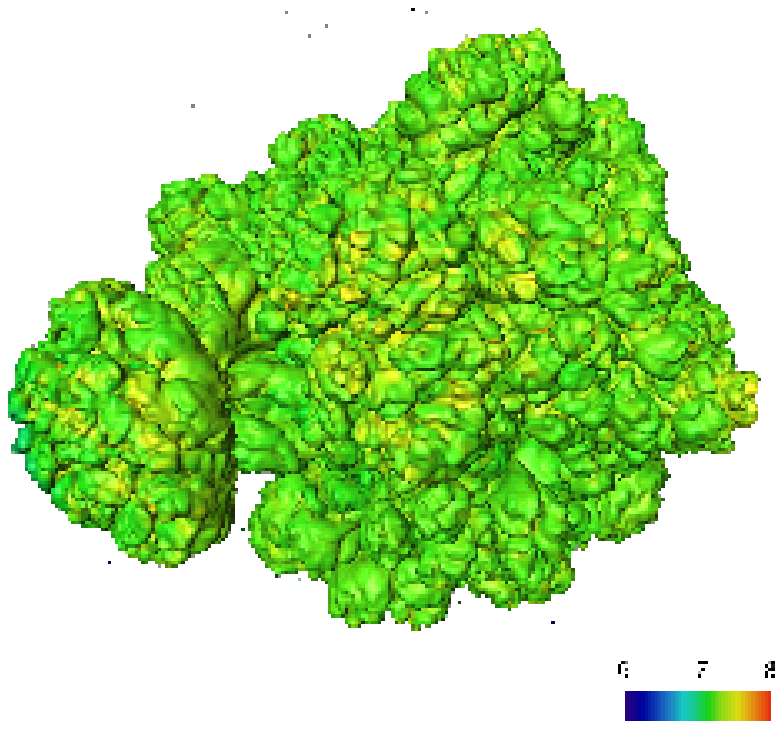} }}
    \caption{Flame evolution in a full-star simulation with contours
      of the subgrid scale turbulence velocity in logarithmic
      scaling. } For each snapshot, the physical time $t$ and the
      number of cells, $N_{\mathrm{uni}}$, as well as the size
      $X_{\mathrm{uni}}$ of the uniform part of the numerical grid
      which encloses the burning region are specified.
    \label{fg:pop_4pi}
  \end{center}
\end{figure*}

For a larger number of ignitions, on the other hand, strong explosions
may result from pure deflagrations. However, even the most energetic
explosions found in our simulations can not account for the upper
range of observed nickel masses. In this respect, we arrive at the same
conclusion as \citet{RoepHille05c} for simultaneous multi-spot
ignitions.  On the other, we found that the explosion ejecta tend to
exhibit an increasingly layered structure with growing number of
ignition events. Consequently, the stochastic ignition model catches
up with simulations in which a delayed detonation is initiated. Apart
from that, \citet{StrizLeib05} have found indications that possibly
more nickel is mixed into the outer layers than previously
assumed. This would confirm the prediction of simulations with an
intermediate number of ignitions, say, about $50$ per octant
(Fig.~\ref{fg:4pi_256ii} and~\ref{fg:pop_4pi}).  Nevertheless, if the
possibility of more than a few ignitions occurring within a tenth of a
second was excluded by advances in understanding the pre-supernova
core evolution, then the deflagration model would be ruled out as an
explanation for typical observed SNe Ia.

\begin{acknowledgements}
We thank Alan Kerstein and Stanford Woosley for helpful discussions
and Friedrich R\"{o}pke for kindly sharing the co-moving grid
implementation. The simulations were run on the HLRB of the Leibniz
Computing Centre in Munich. This work was supported by the Alfried
Krupp Prize for Young University Teachers of the Alfried Krupp von
Bohlen und Halbach Foundation.
\end{acknowledgements}

\bibliographystyle{aa}

\bibliography{StochIgnt_AA}

\end{document}